\documentclass[journal=jpclcd,manuscript=letter]{achemso}

\usepackage{chemformula} 
\usepackage[T1]{fontenc} 
\usepackage{float}
\usepackage{dcolumn}
\usepackage{graphicx}
\usepackage{siunitx}
\usepackage{amsmath}
\usepackage{amssymb}
\usepackage{braket}
\usepackage{float}
\usepackage{graphicx}
\usepackage{siunitx}
\usepackage{amsfonts}
\usepackage{color}
\usepackage{soul}
\usepackage{braket}
\usepackage{verbatim}

\newcommand{\beq}{\begin{equation}}
	\newcommand{\eeq}{\end{equation}}
\newcommand{\E}{\mathcal{E}}

\author{Giovanni Batignani}
\affiliation[Rome]
{Dipartimento di Fisica,~Universit\'a~di~Roma~``La Sapienza",  ~Roma, ~I-00185, ~Italy}
\author{Carino Ferrante}
\affiliation[Rome]
{Dipartimento di Fisica,~Universit\'a~di~Roma~``La Sapienza",  ~Roma, ~I-00185, ~Italy}
\alsoaffiliation[IIT]
{~Istituto~Italiano~di~Tecnologia, Center for Life Nano Science @Sapienza, Roma, ~I-00161,  ~Italy}
\author{Tullio Scopigno}
\affiliation[Rome]
{Dipartimento di Fisica,~Universit\'a~di~Roma~``La Sapienza",  ~Roma, ~I-00185, ~Italy}
\alsoaffiliation[IIT]
{~Istituto~Italiano~di~Tecnologia, Center for Life Nano Science @Sapienza, Roma, ~I-00161,  ~Italy}
\email{tullio.scopigno@roma1.infn.it}

\title{Accessing Excited State Molecular Vibrations by Femtosecond Stimulated Raman Spectroscopy}
\abbreviations{ISRS}

\begin{document}
	
		\begin{tocentry}
		
		\includegraphics[width=1\textwidth]{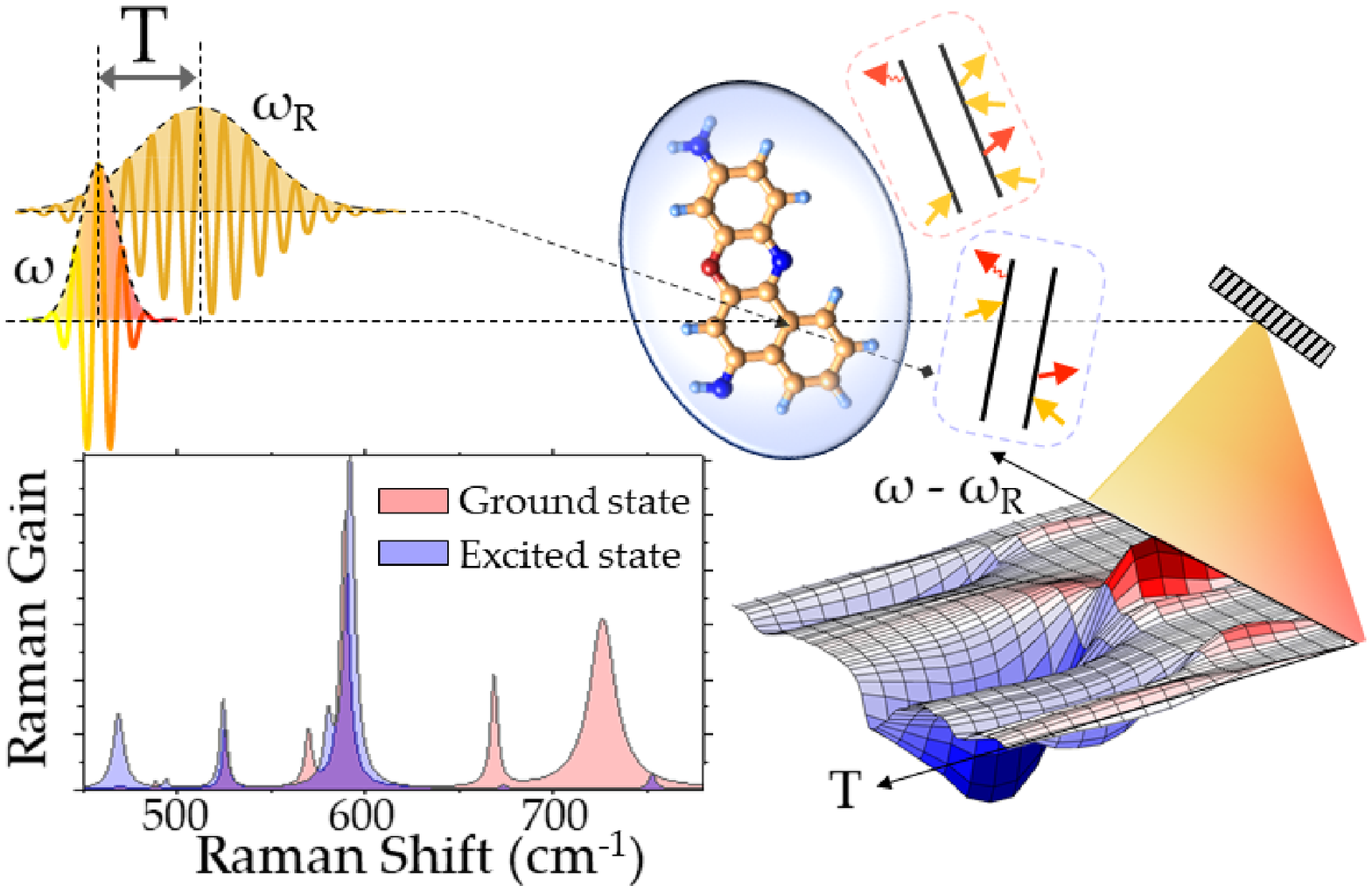}
		
	\end{tocentry}

	\begin{abstract}
		Excited-state vibrations are crucial for determining photophysical and photochemical properties of molecular compounds. Stimulated Raman scattering can coherently stimulate and probe molecular vibrations with optical pulses, but it is generally restricted to ground state properties. Working in resonance conditions, indeed, enables cross-section enhancement and selective excitation to a targeted electronic level, but is hampered by an increased signal complexity due to the presence of overlapping spectral contributions. Here, we show how detailed information on ground and excited state vibrations can be disentangled, by exploiting the relative time delay between Raman and probe pulses to control the excited state population, combined with a diagrammatic formalism to dissect the pathways concurring to the signal generation. The proposed method is then exploited to elucidate the vibrational properties of ground and excited electronic states in the paradigmatic case of Cresyl Violet. We anticipate that the presented approach holds the potential for selective mapping the reaction coordinates pertaining to transient electronic stages implied in photo-active compounds.
	\end{abstract}
\maketitle

Raman spectroscopy is a powerful tool to access the vibrational fingerprints of molecules or solid state compounds and it can be used to extract structural and dynamical information of the samples under investigation. During the last decades, thanks to the development of ultrafast and nonlinear optical techniques~\cite{Zewail2000}, different experimental protocols based on coherent Raman scattering~\cite{Mukamel_Potma_CRS} have been developed for investigating the vibrational properties of reacting species, as well as for studying their photophysical and photochemical properties. 
Particular efforts have been devoted to  the development of experimental and theoretical protocols able to measure and assign vibrational modes on excited potential energy surfaces, distinguishing them from ground state eigenstates~\cite{Bardeen1995,cit::RuhmanWand,Biggs2013X,cit::pcivs::kukura,Dorfman2015X,Dorfman2016,Versteeg2018,Maiuri_2018,Kuramochi_2019,Fumero2020}.

Femtosecond Stimulated Raman scattering (FSRS) represents a convenient way to combine the vibrational sensitivity of Raman spectroscopy with the efficiency of coherent process based techniques, providing high intensity and fluorescence background free signals~\cite{cit::Mukamel}. FSRS exploits the combination of a narrowband Raman pulse (RP) with a femtosecond probe pulse (PP) to coherently stimulate and probe vibrational excitations, read out as Raman gain on the high directional PP field and probed over a wide spectral range. 
Hence, 
the stimulated Raman spectrum can be conveniently isolated considering the ratio $RG=\frac{I_{{On}}(\omega)}{I_{{Off}}(\omega)}$ between the PP spectrum measured in presence ($I_{{On}}$) and in absence ($I_{{Off}}$) of the RP, conventionally referred to as the Raman Gain, providing high structural sensitivity~\cite{Freudiger1857,Zong2019,DelaCadena2020} on the sample under investigation. 
The use of a broadband probe pulse provides the chance to access the entire vibrational spectrum, from low to high frequency Raman modes, in a single acquisition. Additionally, FSRS is immune to non vibrationally resonant background, which on the contrary can overwhelm the vibrational response measured by different frequency domain nonlinear Raman experimental layouts, such as coherent anti-Stokes Raman scattering, both in off-resonant and resonant regimes~\cite{Mukamel_Potma_CRS,Nestor1976,Virga2019}.   
Moreover, since FSRS is able to combine a high spectral resolution with a femtosecond time precision in the stimulation of Raman coherences~\cite{Biggs2011,Fumero2015}, adding a third pulse, namely the photochemical pump that precedes the RP-PP pair and excites the sample, turns FSRS into a pump-probe technique~\cite{Yoshizawa1994,Yoshizawa2000,Mathies_review} able 
to  access vibrational spectroscopy on sub-picosecond time regimes, ensuring at the same time atomic spectral resolution~\cite{Kukura2005,Laimgruber2006,Fang2009,Provencher2014,Zhou2015,Batignani2016,Ferrante2016,Barclay2017,Hall2017,Piontkowski2018,Quincy2018,Barclay2019,Batignani2019_adp, Sotome2020}.   

Notably, Raman features in the FSRS scheme provide information on the imaginary part of the generated nonlinear polarization, due to the self-heterodyne nature of the measured signal ~\cite{cit::Mukamel,Mukamel_Potma_CRS,Prince2016}.
Hence the spectral profiles can result in complex lineshapes, depending on the probed spectral range, on the RP resonance condition and on the order of the radiation-matter interactions that generates the nonlinear Raman signal~\cite{cit::Mukamel,Batignani2016SciRep}.
Specifically, for a RP wavelength $\lambda_{RP}$ tuned far from any electronic transition of the system, FSRS bands appearing to the red side of the spectrum (at PP wavelengths $\lambda$ larger than $\lambda_{RP}$) are always positive gains, while FSRS features to the blue side are negative losses~\cite{Mukamel_2013}. 
On the contrary, for a resonantly tuned RP, the FSRS signals appearing to the blue side of the spectrum show a profile evolving from negative losses to positive gains through dispersive lineshapes as a function of the resonance condition~\cite{Batignani2016SciRep}, while FSRS bands to the red side typically show positive profiles.
Critically, in presence of a resonant RP, Raman coherences can be generated on different potential energy surfaces, therefore assigning  the measured  vibrations  to the pertaining electronic state can represent a demanding task. 
In particular, 
while the resonance enhancement is in general a powerful tool for isolating the vibrational response of a desired chromophore in complex molecular systems~\cite{Spiro1975},
in presence of overlapping ground state absorption and stimulated emission, it is ineffective for discriminating between signals originated from ground or excited energy levels. 
This has so far limited the use of FSRS for mapping excited state properties with respect to its time-domain analogues, namely impulsive Stimulated Raman Scattering based techniques\cite{Liebel2013,Fujisawa2016,Rafiq2016,Park2018,Kuramochi2019,Kim2020}, where the coherent oscillations can be directly monitored in the time-domain. In particular, phase analysis or dependence of the time-domain signal on the pulse chirp  are a basis for distinguishing between ground or excited electronic state vibrations\cite{Zeiger1992,Monacelli2017,Batignani2018}. 

\begin{figure}[h!]
	\centerline{\includegraphics[width=18cm]{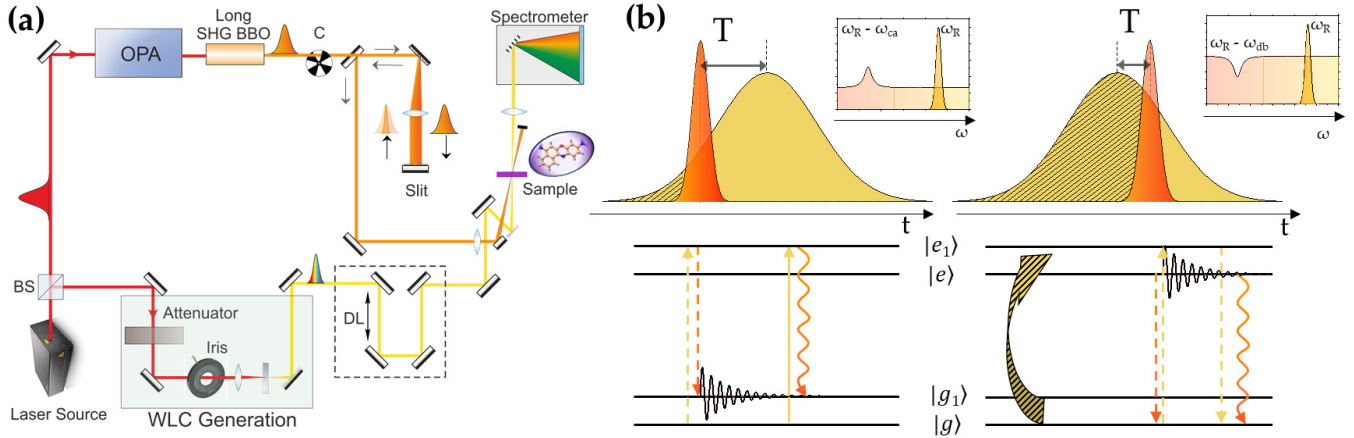}}
	\caption{Two-pulse FSRS concept in presence of a resonant Raman pulse: (a) sketch of the experimental setup and (b) pulse interactions scheme.
	For a PP preceding the RP maximum vibrational coherences are stimulated in the ground electronic state and result in positive gains on the PP spectrum. On the contrary, for a RP preceding the PP a large portion of the Raman pulse can be absorbed by the system, and excited state properties can be probed by subsequent FSRS  interactions with the RP-PP pair. 
	In the bottom of panel (b) we report energy ladder diagrams describing the two different processes. $\ket{g}$ and $\ket{g_1}$ denote the ground and the first vibrational excited levels in the electronic ground state, while $\ket{e}$ and $\ket{e_1}$ indicate their counterpart in the excited electronic level.
	BBO: Beta Barium Borate; BS: Beam Splitter; DL: Delay Line; OPA: Optical Parametric Amplifier; SHG: Second Harmonic Generation; 
	 \label{Fig: Sketch}}
\end{figure}

In this work we consider an FSRS experiment where a high fluence resonant Raman pulse, which, acting also as a photochemical pump, promotes the system ground state population to a desired electronic state previous to the coherent stimulation of the vibrational coherences. 
Interestingly, FSRS experiments are commonly performed using a positive time delay $T$ between the Raman and the probe pulses (i.e. with the PP that temporally precedes the RP maximum), in order to increase both the Raman gain and the spectral resolution~\cite{Yoon2005,Ferrante2018}. In fact, as depicted in the left panels of Fig.~\ref{Fig: Sketch}, since the stimulated vibrational coherences evolve until a second interaction with the RP that gates the Raman oscillations, using a positive $T$ enables increasing the effective temporal window in which the Raman coherences are sampled. Under such condition, only a small portion of the RP precedes the PP and hence most of the molecules are in the ground state at the arrival time of the probe.
Here we demonstrate how the opposite scenario, i.e. a PP temporally following the RP maximum, can be exploited to probe excited state vibrational properties by photoexciting the system in a controlled manner acting on the negative time delay, and then stimulating the Raman process thanks to the joint action of the residual RP with the PP. 
\begin{figure*}
	\centerline{\includegraphics[width=16cm]{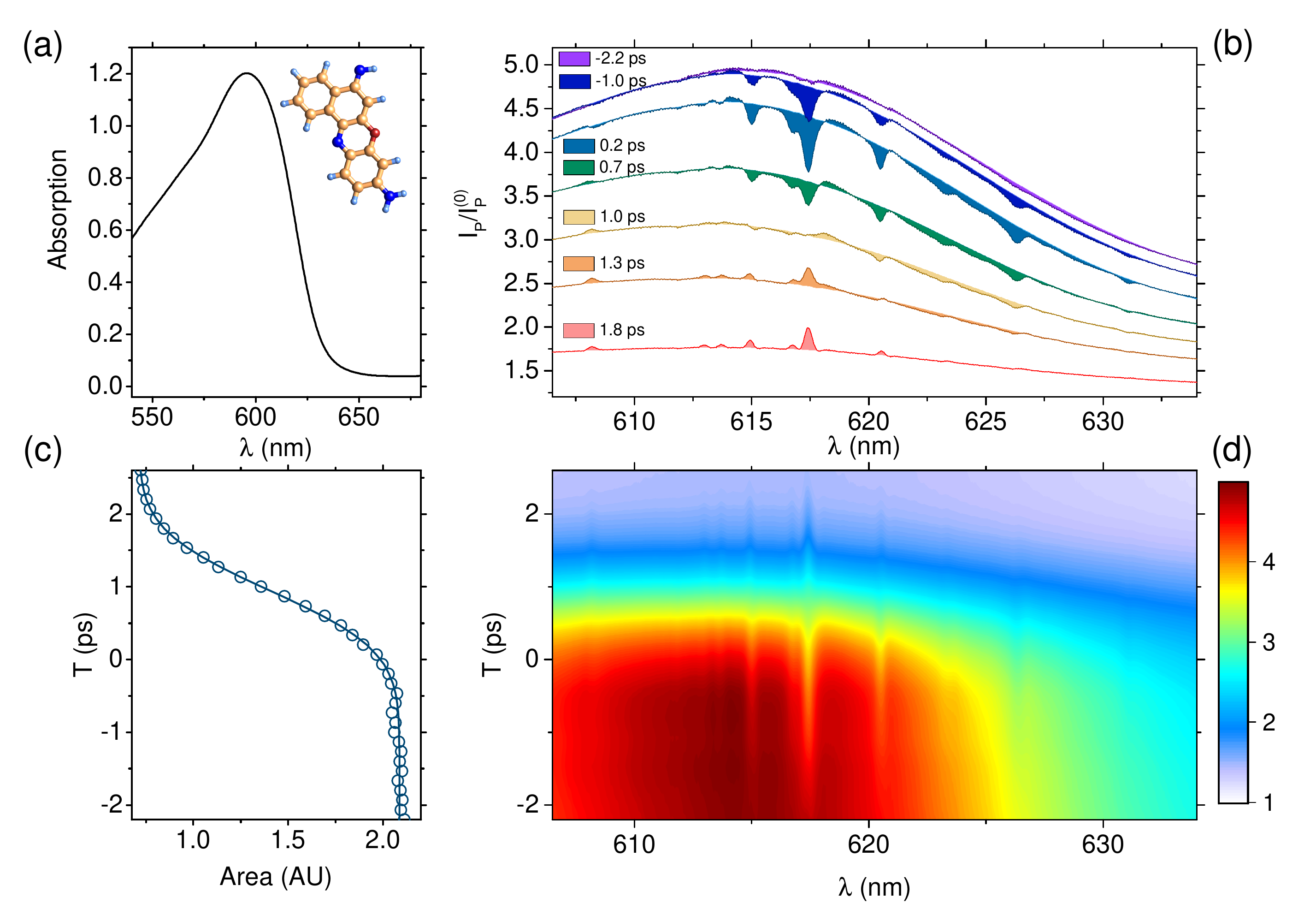}}
	\caption{(a) Cresyl Violet static absorption spectrum, with a sketch of the sample molecular structure.
		Due to the RP-induced photo-excitation, FSRS spectra are accompanied by a modification of the PP absorption, which results in a baseline background superimposed to the Raman features.
		In panel (b) we report the FSRS spectrum recorded with a 420 nJ Raman pump tuned at $\sim$ 595 nm  before the baseline removal and shown for selected time delays T between the Raman and the probe pulses; the filled areas represent the FSRS	isolated contributions. The baseline maximum is red-shifted with respect to the maximum absorption due to a small ($\approx$ 20 nm) Stokes shift.
		We note that for a PP following the RP the baseline intensity decreases, indicating a reduced quantity of molecules in the electronic excited state (different time delay traces have not been vertically offset). While for positive time delays Raman bands appear as peaks on the top of baseline background, for negative T the FSRS spectrum shows negative peaks (losses).
		In panel (d) the corresponding color maps are reported for all of the measured time delays, while in (c) we show the baseline area as a function of T, which can be used to extract a direct estimate of the excited vs ground  states molecules.\label{Fig: Absorption_Bsl_SRS}}
\end{figure*}
In order to evaluate the femtosecond stimulated Raman response under such conditions, we recorded red side FSRS spectra of cresyl violet (CV)~\cite{Isak1992,Vogel2000,Brazard2015,Rafiq2016}, an oxazine dye commonly used in histology as stain, characterized by a long living excited state,  with the non-radiative decay to the ground state occurring on the nanosecond timescale. Moreover, CV shows overlapping ground state absorption and stimulated emission, representing hence an excellent candidate for testing the capability of the presented approach for distinguishing between ground and excited state vibrations.
CV was dissolved in methanol and the FSRS spectrum was measured at ambient temperature with the RP tuned to be in resonance with the sample absorption maximum at $\sim$ 595 nm. 
A sketch of the two-pulse FSRS experimental setup exploited for the measurements is reported in Fig. \ref{Fig: Sketch}(a). Briefly, a Ti:sapphire laser source that generates 3.6 mJ, 35 fs pulses at 800 nm and 1 kHz repetition rate. The synthesis of the RP is obtained by a two-stage optical parametric amplifier (OPA) that produces tunable near-infrared pulses, followed by a frequency doubling inside a 25 mm thick Beta Barium Borate (BBO) crystal. 
Spectral compression inside the thick nonlinear crystal ensures at the same time high RP fluences and narrowband pulses~\cite{Marangoni_Cerullo_picosecond}. An additional rectification of the RP temporal and spectral profiles is achieved by a doublepass (2f) spectral filter~\cite{Pontecorvo_2013,Hoffman2013}, which results in Gaussian $\sim$ 2.5 ps Raman pulses. The femtosecond PP is a white-light continuum (WLC) obtained by focusing a small portion of the laser fundamental into a Sapphire crystal. A variable attenuator and an iris along the beam path before the Sapphire crystal are used to tune and optimize the shape of the generated WLC. RP polarization is parallel to the PP and its fluence is adjusted by using a linear neutral density filer. Further details on the experimental scheme have been described in Refs. \cite{Batignani2019,Ferrante2020}.

The absorption spectrum of the system, together with a sketch of the molecular structure, is reported in Fig.~\ref{Fig: Absorption_Bsl_SRS}a, while in Fig.~\ref{Fig: Absorption_Bsl_SRS}b we report the measured Raman spectra for selected time delays $T$, with the corresponding FSRS color maps for all the measured $T$ shown in panel d. 
Notably, the RP photo-excitation induces a modification of the PP transmission, which results in a smooth background due to ground state bleaching and stimulated emission additional to the FSRS features. The area under such a baseline (shown in Fig.~\ref{Fig: Absorption_Bsl_SRS}c) conveniently provides a direct estimate of the $T$-dependent fraction of molecules promoted to the excited electronic state, and can be determined by subtracting a polynomial profile obtained as best fit to the Raman gain in the spectral regions free from Raman bands. As expected, for large positive time delays the baseline vanishes, indicating that all the molecules probed by FSRS are in the ground state, while for lower $T$ values the baseline area increases reaching an almost constant value for $T<$ -500 fs.  
Most importantly, the FSRS lineshapes significantly evolve with the delay: the usual positive Raman gains observed for large $T$ become negative bands upon reducing $T$ with a trend following the baseline area, suggesting FSRS contributions from excited state vibrational coherences.

To get further insights on the origin of such negative profiles, we modeled the measured spectra by evaluating FSRS response trough a perturbative expansion of the density matrix in powers of the $E(t)=\sum_{i} \E_{i}(t) e^{-i\omega_i t} +c.c.$ electric fields~\cite{cit::Mukamel,Batignani2015pccp,Agarwalla2015}. 
In particular, the Raman Gain can be calculated as
\begin{equation}\label{Eq: Raman Gain}
RG(\omega)=-\Im\left[\frac{P^{(n)}(\omega,T,\lambda_R)}{\E_S^{0}(\omega)}\right]
\end{equation}
where $\Im(x)$ indicates the imaginary part of $x$, $\E_S^{0}(\omega)$ is the probe field spectral profile (before interacting with the sample) and $P^{(n)}(\omega,T,\lambda_R)$ is the $n^{\hbox{th}}$ order total nonlinear polarization induced in the system~\cite{Dorfman2013}.
Since, several processes, corresponding to different RP and PP field permutations, contribute to the generation of the FSRS signal, several Feynman diagrams, depicting the density matrix evolution during consecutive interactions with the electromagnetic fields, have to be considered. 

We evaluated the third order FSRS process with the system initially either in the ground state ($\ket{g}\bra{g}$) or starting from the excited electronic level ($\ket{e}\bra{e}$) upon two preceding interactions with the RP. The initial excited state population,  proportional to the measured baseline area, can be directly included in the model from the experimental traces. 
\begin{figure*}[ht]
	\centerline{\includegraphics[width=11cm]{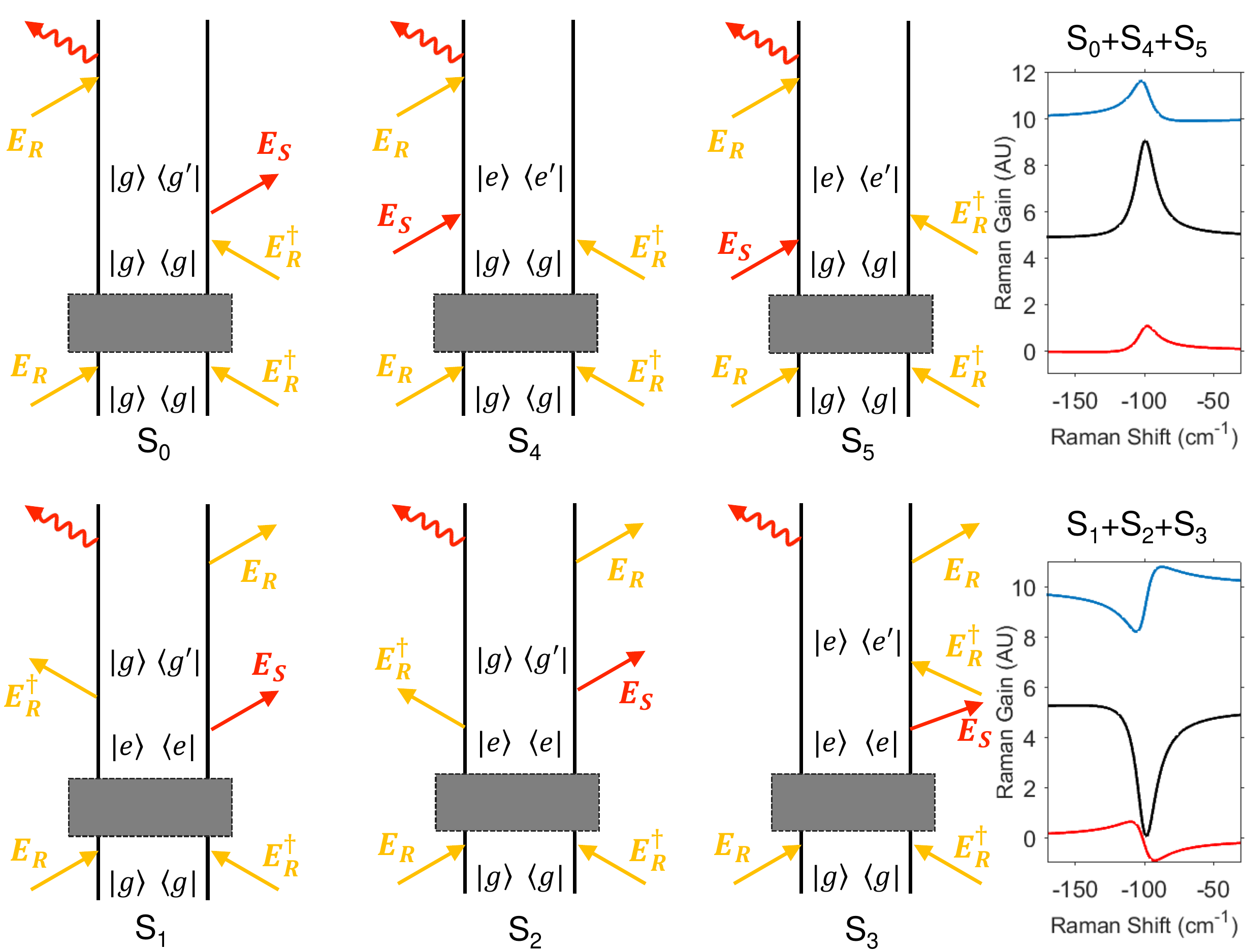}}
	\caption{Feynman diagrams describing the FSRS process upon a RP induced excitation to the $e$ electronic state. Upon a double interaction with the RP, the system can be either in the ground or in the excited electronic states ($\ket{g}\bra{g}$ and $\ket{e}\bra{e}$ populations, respectively). 
	Upon the joint action of the RP-PP pair, a vibrational coherence, either in the ground ($\ket{g}\bra{g'}$) or in the excited electronic state ($\ket{e}\bra{e'}$), is stimulated.	
	In the right panels, we evaluate the $S_0+S_4+5$ and $S_1+S_2+3$ signals for a model system with a Raman active eigenstate at 100 cm$^{-1}$ mode, considering a monochromatic RP with a wavelength matching the electronic transition and equal dipole moments (black lines): while the system response associate to processes starting from the ground state ($S_0+S_4+5$ diagrams) is a positive gain, diagrams associated to a system initially in the electronic excited state ($S_1+S_2+3$) result in a negative loss. Red and blue lines show the calculated $S_0+S_4+5$ and $S_1+S_2+3$ signals obtained partially detuning the RP wavelength (considering a 16 nm shift). The traces corresponding to a different resonance condition have been vertically offset by a constant factor.
		\label{Fig: F_diagrams}}
\end{figure*}
The Feynman diagrams that take into account for the FSRS response are shown in Fig. \ref{Fig: F_diagrams}. In the top panel we report the $S_0$, $S_4$ and $S_5$ processes, where the first RP-PP interactions act on the system initially in the electronic ground state ($\ket{g}\bra{g}$ population), while in the bottom panel the diagrams $S_1$, $S_2$ and $S_3$ are associated to processes where a double interaction with the RP initially promotes the system to the excited electronic state $\ket{e}\bra{e}$, which is then interrogated by a subsequent third order FSRS process.
The nonlinear polarization associated to $S_0$ can be expressed as
\begin{multline}\label{Eq: P3t_S0}
P^{(3)}_{S0}(t,T)=
\sum_{g'e}
\left(\frac{i}{\hbar}\right)^3
P_{gg} |\mu_{ge}|^2|\mu_{eg'}|^2
\int_0^{\infty}\int_0^{\infty}\int_0^{\infty}\\
d\tau_3d\tau_2d\tau_1\E_R^*(t-\tau_1-\tau_2-\tau_3,T)\E_S(t-\tau_2 -\tau_3)\E_R(t-\tau_3,T)
\\
e^{i\omega_R (-\tau_1-\tau_2)}
e^{-i\omega_S (t-\tau_2-\tau_3)}
e^{-i\tilde{\omega}_{ge} \tau_1}
e^{-i\tilde{\omega}_{gg'} \tau_2}
e^{-i\tilde{\omega}_{eg'} \tau_3}
\end{multline}
where $T$ is the relative time delay between Raman and probe pulses, the summation over $g'$ runs over all the vibrationally excited states in the ground electronic level and $e$ over the states in the excited electronic level (see Fig. \ref{Fig: F_diagrams}),  $\tilde{\omega}_{ij}=\omega_{ij} -i\gamma_{ij}$. Here, $\omega_{ij}=\omega_i-\omega_j$, $\gamma_{ij}$ indicates the dephasing rate of the
$\ket{i}\bra{j}$ coherence, $\mu_{ij}$ is the dipole transition moment between $i$ and $j$ states and 
$P_{gg}$ is the ground state population at the arrival time of the PP. 
By evaluating the system response in the frequency domain and writing the pulse fields in terms of their Fourier transforms \cite{Dorfman2013,Batignani2019_CISRS}
$
\E_{S/R}(t)=\int_{-\infty}^{\infty}d\omega \E_{S/R}(\omega)e^{-i\omega t}
$
we obtain
\begin{equation}
P^{(3)}_{S0}(\omega,T)=-
\sum_{g'e}
\int_{-\infty}^{\infty}
\int_{-\infty}^{\infty}d\omega_1 d\omega_3
\frac{P_{gg}|\mu_{ge}|^2|\mu_{eg'}|^2
	\E_R^*(\omega_1,T)\E_S(\omega+\omega_1-\omega_S-\omega_3)\E_R(\omega_3,T)}
{	\hbar^3(\omega_R+\omega_1+\tilde{\omega}_{ge})
	(\omega_R-\omega+\omega_3+\tilde{\omega}_{gg'})
	(\omega-\tilde{\omega}_{eg'})}
\end{equation}
Similarly, the other nonlinear polarization terms associated to diagrams $S_1$--$S_5$ can be obtained and expressed as
\begin{equation}
P^{(3)}_{S1}(\omega,T)=
\sum_{g'}
\int_{-\infty}^{\infty}
\int_{-\infty}^{\infty}
d\omega_2 d\omega_3
\frac{
P_{ee}	|\mu_{eg'}|^2|\mu_{ge}|^2
	\E_S(\omega-\omega_S+\omega_2-\omega_3)\E_R^*(\omega_2,T)\E_R(\omega_3,T)}
{\hbar^3(\omega+\omega_2-\omega_3-\tilde{\omega}_{eg'})
	(\omega-\omega_3-\omega_R-\tilde{\omega}_{gg'})
	(\omega-\tilde{\omega}_{eg'})}
\end{equation}
\begin{equation}
P^{(3)}_{S2}(\omega,T)=
\sum_{g'}
\int_{-\infty}^{\infty}
\int_{-\infty}^{\infty}d\omega_1d\omega_3
\frac{P_{ee}|\mu_{eg'}|^2|\mu_{ge}|^2
	\E_R^*(\omega_1,T)
	\E_S(\omega_+\omega_1-\omega_S-\omega_3)
	\E_R(\omega_3,T)}
{	\hbar^3(\omega_R+\omega_1+\tilde{\omega}_{ge})
	(\omega_R-\omega+\omega_3+\tilde{\omega}_{gg'})
	(\omega-\tilde{\omega}_{eg'})}
\end{equation}
\begin{equation}\label{eq:P3_S3}
P^{(3)}_{S3}(\omega,T)=
\sum_{e'g'}
\int_{-\infty}^{\infty}
\int_{-\infty}^{\infty}d\omega_2d\omega_3
\frac{P_{ee}|\mu_{g'e}|^2|\mu_{g'e'}|^2
	\E_S(\omega-\omega_S+\omega_2-\omega_3)\E_R^*(\omega_2,T)\E_R(\omega_3,T)}
{\hbar^3(\omega+\omega_2-\omega_3-\tilde{\omega}_{eg'})
	(\omega-\omega_3-\omega_R-\tilde{\omega}_{ee'})
	(\omega-\tilde{\omega}_{eg'})}
\end{equation}
\begin{equation}
P^{(3)}_{S4}(\omega,T)=-
\sum_{e'}
\int_{-\infty}^{\infty}
\int_{-\infty}^{\infty}d\omega_1d\omega_3
\frac{P_{ee}
	|\mu_{ge'}|^2|\mu_{ge}|^2\E_R^*(\omega_1,T)\E_S(\omega+\omega_1-\omega_S-\omega_3)\E_R(\omega_3,T)}
{\hbar^3(\omega_R+\omega_1+\tilde{\omega}_{ge'})
	(\omega_R-\omega+\omega_3+\tilde{\omega}_{ee'})
	(\omega-\tilde{\omega}_{eg})}
\end{equation}
\begin{equation}\label{Eq: P3t_S5}
P^{(3)}_{S5}(\omega,T)=-
\sum_{e'}
\int_{-\infty}^{\infty}
\int_{-\infty}^{\infty}
d\omega_2d\omega_3
\frac{P_{gg}
	|\mu_{ge'}|^2|\mu_{ge}|^2
	\E_S(\omega-\omega_S+\omega_2-\omega_3)\E_R^*(\omega_2,T)\E_R(\omega_3,T)}
{\hbar^3(\omega+\omega_2-\omega_3-\tilde{\omega}_{eg})
	(\omega-\omega_3-\omega_R-\tilde{\omega}_{ee'})
	(\omega-\tilde{\omega}_{eg})}
\end{equation}
where $P_{ee}$ indicates the electronic excited state population at the arrival time of the PP. A complete derivation of the $S_0$-$S_5$ signals is reported in the Supporting Information. We note that in order to correctly evaluate the total system response in presence of a resonant RP, diagrams $S_4$ and $S_5$, which consider a system initially in the ground state, with a vibrational coherence $\ket{e}\bra{e'}$ generated upon the joint action of the RP-PP pair, should be considered. Interestingly, as clarified in the Supporting Information, for an off-resonant Raman pulse, the $S_4$ and $S_5$ responses result in a Lorentzian response with opposite sign and hence cancel out.

\begin{figure*}
	\centerline{\includegraphics[width=19cm]{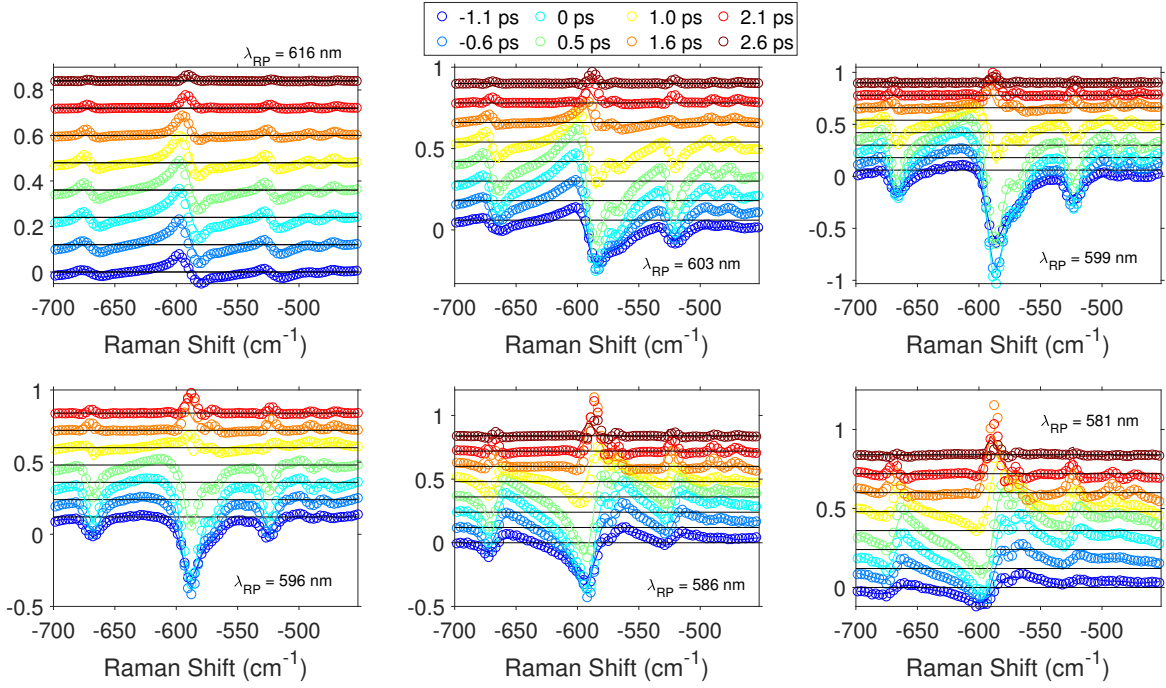}}
	\caption{Wavelength dependence of the FSRS spectra: data (circles) and model (continuous lines) are compared as a function of RP wavelength $\lambda_{RP}$ and relative time delay $T$ between the Raman and the probe pulses. For a RP tuned to be in resonance with a system electronic transition, Raman excitation processes arising from the system initially in the excited state result in intense negative losses (negative time delays, with a RP at 596-599 nm). Tuning the RP away from a full resonant condition turns the negative signal into a dispersive profile, with an odd symmetry: blue-shifting the RP the positive lobe is at higher wavenumbers, a red-shift of the RP generates a positive lobe at lower frequencies. Traces acquired at different $T$ have been vertically offset by a constant factor.\label{Fig: Fit_Slices}}
\end{figure*}

In the right panel of Fig. \ref{Fig: F_diagrams} we show the corresponding $S_0$--$S_5$ signals, with $S_j=-\Im\left[\frac{P^{(3)}_{S_j}(\omega)}{\E_S^{0}(\omega)}\right]$, evaluated  considering for simplicity equal weights and dipole moments for all the diagrams: for a perfectly resonant RP,
while the system response associated to processes starting from $\ket{g}\bra{g}$ is a positive gain, diagrams where the system is initially in an excited electronic state population result in a negative loss. 
Notably, detuning the RP away from a perfectly resonant condition is almost ineffective on the Raman lineshape of $S_0$, $S_4$, $S_5$ diagrams and only reduces their corresponding signal intensity. On the other hand, it results in a dispersive profile for $S_1$, $S_2$, $S_3$, i.e. for those pathways starting from the excited  $\ket{e}\bra{e}$ state. 
As shown in the right panel of Fig. \ref{Fig: F_diagrams}, for a RP red-shifted with respect to the maximum absorption, the positive lobe of such dispersive profile is at lower frequencies (higher absolute Raman Shift), while blue-shifting the RP generates a positive lobe at lower frequencies.
Interestingly, for the diagram $S_3$, which involves the creation of excited state vibrational coherences, all the first three interaction occur on the bra side of the density matrix, resulting in a mode specific resonant enhancement condition similar to the one characterizing the ground state blue side SRS response \cite{Batignani2016SciRep}.
In fact, the first and the third denominator in Eq. \ref{eq:P3_S3} give rise to a resonant enhancement for a RP wavelength matching the energy difference $\hbar\left(\omega_{eg}+\omega_{e'e}\right)$.
\begin{figure}[h]
	\centerline{\includegraphics[width=9cm]{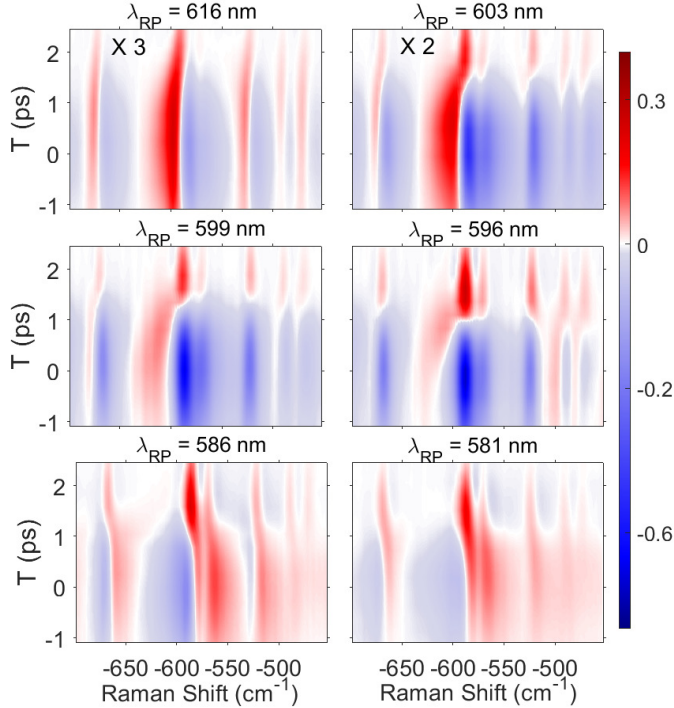}}
	\caption{RP wavelength dependence of the FSRS spectra as a function of wavenumber and relative time delay between RP and PP. While for positive time delays, where the signal is originated from processes starting from the ground state, only positive Raman gains are recorded, for negative T we observe a signal profile that depends on the RP wavelength, as expected in presence of signal involving excited electronic state excitations. 
		\label{Fig: DataMAP}}
\end{figure}
In order to verify the presence of such dispersive profiles  in the measured spectra and hence also the presence of $\ket{e}\bra{e}$ excitations, we collected cresyl violet FSRS spectra
scanning the RP wavelength across the resonance profile.  In  Fig. \ref{Fig: DataMAP} we report the corresponding colormap  as a function of Raman shift and temporal delay T between the RP-PP pair. 
Slices for selected time delays are also reported in Fig. \ref{Fig: Fit_Slices} as colored circles.
As expected, while for high positive T the RP--PP interaction stimulating the vibrational coherences only acts on those molecules initially in the ground state and generates only positive Raman gains, for negative T we observe negative or dispersive lineshapes indicating the presence of excited state Raman coherences. 
We note that also for some positive time delays (T $<$ 1 ps), a significant fraction of molecules is in the excited state at the arrival time of the probe

\begin{figure}[h]
	\centerline{\includegraphics[width=9cm]{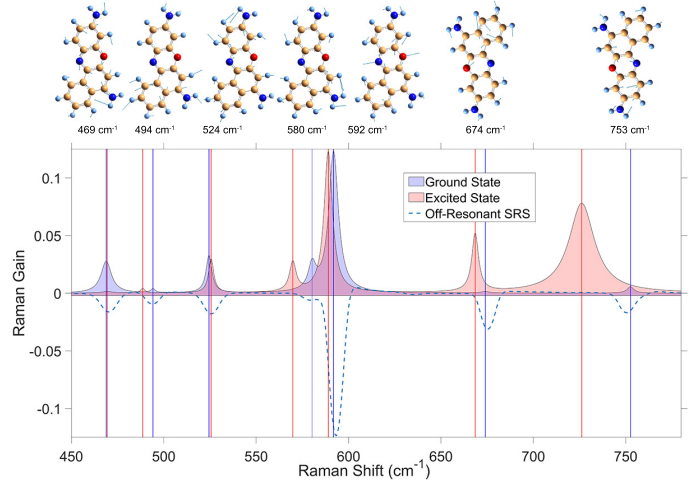}}
	\caption{Comparison between the non resonant FSRS spectrum recorded in the blue side (dashed line) and the extracted resonant Raman profiles for ground and excited electronic states coherences (blue and red filled areas, respectively), with the Raman Gains that have been normalized at the intensity of the stronger band (at $\approx$ 590 cm${-1}$). Vertical blue and red lines indicate the extracted  Raman mode frequencies in the ground and in the excited states, respectively. 
	Extracted and measured ground state peak central positions are in agreement, while the  different Raman intensities are expected in view of the resonant vs off resonant conditions. The presence of purely negative peaks ensures the complete off-resonant condition for the blue side FSRS spectrum \cite{Batignani2016SciRep}.
	In the top panel sketches of the ground state elongation eigenvectors computed by density functional theory with B3LYP functional and the  6-311++G(d,p) basis set are shown\cite{Becke1988,Lee1988}.
		\label{Fig: Blue_Vs_simulated}}
\end{figure}

In order to retrieve the vibrational information associated to the measured FSRS spectra, 
we performed a global fit on the experimental traces using Eqs. \ref{Eq: Raman Gain}-\ref{Eq: P3t_S5}.
The Raman pulse $\E_R(t)$ and the probe pulse $\E_S(t)$ temporal envelopes have been modeled as Gaussian profiles:
$$
\E_{R}(t)=\E_{R}^0 e^{-\frac{(t-T)^2}{2\sigma_{R}^2}}e^{-i\omega_{R} t}
$$
$$
\E_{S}(t)=\E_{S}^0 e^{-\frac{t^2}{2\sigma_{P}^2}}e^{-i\omega_{S} t}
$$
where the parameters $\omega_{ba}$ and $\sigma_{R}$ have been adjusted to best fit the experimental data ($\omega_{ba} = 16900$ cm$^{-1}$, the electronic dephasing rate $\gamma_{ba} = 580$ cm$^{-1}$, with an extracted RP duration equal to $\approx$ 2.5 ps in agreement with the experimental one), while the initial ground and excited state populations $P_{gg}$ and $P_{ee}$ have been extracted from the baseline area.
In Fig. \ref{Fig: Fit_Slices}, the experimental FSRS spectra (colored circles) are compared with the simulations (continuous lines), showing a good agreement. This is further corroborated by the comparison between the extracted ground and excited electronic state FSRS spectrum and the non resonant blue side stimulated Raman spectrum reported in Fig. \ref{Fig: Blue_Vs_simulated}, with peak positions and amplitudes reported in Table \ref{table: FSRS}. As expected, the extracted electronic ground state Raman line positions are in line with the blue side Raman losses, validating the capability of assigning the measured vibrations to the corresponding electronic state. Notably, the excited state normal modes show in general small frequency decreases indicating reduced excited energy level force constants (with respect to the ground state potentials). For the $590$ cm$^{-1}$, $669$  cm$^{-1}$ and $726$ cm$^{-1}$  Raman modes, we identify high transition dipole moments $\mu_{ge'}$, pointing to the reaction coordinate nature of such Raman excitations \cite{cit::Mukamel,Fumero2020}.
\begin{table}
	\begin{tabular}{|c|c||c|c| }
		\hline
			$\tilde{\nu}_{g'g}$ (cm$^{-1}$)  & $\tilde{\nu}_{e'e}$   (cm$^{-1}$)   & $|\mu_{eg'}|^2$ (AU)& $|\mu_{ge'}|^2$ (AU) \\
		\hline
		 469 & 469  &0.7 & 0.1\\
		\hline
		 494 & 489  &0.1 & 0.3\\
		\hline
		  524 & 526  & 0.9 & 2.4\\
		\hline
		 580 & 570 &0.6 & 2.1\\
		\hline
		 592 & 589  &3.3 & 10\\
		\hline
		 674 & 669 &0.04 & 4.2\\
		\hline
		753 & 726 &0.15 & 6.5\\
		\hline
	\end{tabular}
	\caption{Ground and excited states peak positions and intensities.}\label{table: FSRS}
\end{table}    

It is worth to stress that, since several processes concur to the generation of the measured signal and since also a small detuning of the RP from a perfectly resonant condition results in disperive Raman lineshapes, it is crucial to build on Eqs. \ref{Eq: Raman Gain}-\ref{Eq: P3t_S5} for correctly extracting the exact Raman excitation frequencies and linewidths.
This procedure establishes a convenient protocol to measure and assign vibrational properties to a targeted excited electronic state and it is particularly convenient for the challenging case of overlapping ground state absorption and stimulated emission from the excited electronic state, when the conventional resonance enhancement is not effective for discriminating between signals originated from different potential energy surfaces. 

The present results also indicate that large Raman pump fluences do not help, in general, to record more accurate ground state FSRS spectra. Despite the linear dependence of the Raman cross section on the RP intensity, indeed, increasing the RP energy can promote the system to the excited state. Accordingly, concurring signals are generated with dispersive profile or even negative sign, overwhelming the desired Raman information. 
This result rationalizes also why, under resonance condition, the FSRS spectrum measured in the blue side can show more intense Raman bands, providing better signal to noise ratio. In fact, while in the blue side the resonance condition is red-shifted with respect to electronic absorption peak \cite{Batignani2016SciRep} and hence the promotion of the system to the excited state is avoided, in the red side the resonance enhancement is achieved for a Raman wavelength that matches the electronic transition and that can promote the system to the excited state, with the signals originated from excited state coherences that can destructively interfere with the ground state  FSRS response.

In summary, we have investigated the FSRS response in presence of a RP photo-induced excited state population. 
A diagrammatic treatment of the signal generation has been exploited for analyzing the data, scrutinizing the concurring pathways that generate the nonlinear Raman response and discriminating between processes involving ground and excited state coherences.
We have shown how to experimentally control the excited state population tuning the relative time delay between Raman and probe pulses and, by comparing the FSRS response as a function of the Raman resonance condition, we have demonstrated how to extract the excited state vibrational lineshapes, developing a novel FSRS protocol for recording excited state Raman spectra. 
As a benchmark of the proposed  experimental scheme and theoretical model, we applied the technique to study Cresyl violet, dissecting its ground and excited state vibrational properties and identifying reduced force constants in the excited state.  We anticipate that the presented approach holds the potential for selective mapping the reaction coordinates pertaining to different electronic stages implied in photo-active compounds.

\bibliography{biblio3}

\begin{acknowledgement}
G.B. and T.S. acknowledge the `Progetti di Ricerca Medi 2019' grant by Sapienza Universit\'a~di~Roma. 
This project has received funding from the PRIN 2017 Project, Grant No. 201795SBA3-HARVEST, and from the European Union's Horizon 2020 research and innovation program Graphene Flagship under Grant Agreement No. 881603.
\end{acknowledgement}

\end{document}